\documentclass[fleqn,10pt]{wlscirep}
\usepackage[utf8]{inputenc}
\usepackage[T1]{fontenc}

\usepackage{lineno,hyperref}
\usepackage{amsmath,graphicx,makecell}
\usepackage{enumitem}
\graphicspath{{./fig/}}
\usepackage{threeparttable}
\usepackage{booktabs}
\usepackage{algorithmic}
\usepackage{subcaption}
\usepackage[linesnumbered, ruled, vlined]{algorithm2e}
\usepackage{todonotes}   % Added by Daniel for the \todo command, feel free to delete when these are all removed
\usepackage{soul}        % Added by Daniel for the \st command, feel free to delete when these are all removed

\title{The Impact of Heavy-Duty Vehicle Electrification on Large Power Grids: a Synthetic Texas Case Study}

\author[1]{Rayan~El~Helou}
\author[1]{S.~Sivaranjani}
\author[1]{Dileep~Kalathil}
\author[2]{Andrew~Schaper}
\author[1,3,*]{Le~Xie}

\affil[1]{Department of Electrical and Computer Engineering, Texas A\&M University, College Station, Texas, USA}
\affil[2]{Department of Multidisciplinary Engineering, Texas A\&M University, College Station, Texas, USA}
\affil[3]{Texas A\&M Energy Institute, College Station, Texas, USA}
\affil[*]{Corresponding author: le.xie@tamu.edu}

\begin{abstract}
The electrification of heavy-duty vehicles (HDEVs) is a nascent and rapidly emerging avenue for decarbonization of the transportation sector.  %Any mass electrification of such large vehicles requires a substantial restructuring and shift from fuel-based to electricity-based charging infrastructure. This would potentially disrupt both electric grid operators and external market participants, and it is imperative for policy makers to adopt a data-driven, scientific approach to reconciling the market adoption of heavy-duty electric vehicles (HDEVs) with the realities of existing grid infrastructure. 
In this paper, we examine the impacts of increased vehicle electrification on the power grid infrastructure, with particular focus on HDEVs. We utilize a synthetic representation of the 2000-bus Texas transmission grid, and realistic representations of multiple distribution grids in Travis county, Texas, as well as transit data pertaining to HDEVs, to uncover the consequences of HDEV electrification, and expose the limitations imposed by existing electric grid infrastructure. Our analysis reveals that grid-wide voltage problems that are spatiotemporally correlated with the mobility of HDEVs may occur even at modest penetration levels. In fact, we find that as little as 11\% of heavy duty vehicles in Texas charging simultaneously can lead to significant voltage violations on the transmission network that compromise grid reliability. Furthermore,  we find that just a few dozen EVs charging simultaneously can lead to voltage violations at the distribution level.  %can pose a threat to the physical security and reliability of the grid. %Our analysis is generalizable over any set of transmission and distribution networks, and is not restricted to those in Texas, thus revealing the fundamental limitations of the underlying electric grid infrastructure that must be addressed before any serious HDEV electrification effort. 
\vspace{-1.8em}
\end{abstract}

\begin{document}
\flushbottom
\maketitle
\thispagestyle{empty}

% \todo[inline]{Daniel: in the Abstract, does the analysis `uncover' the regional disparity of load shedding, or is that already apparent in the source data from ERCOT?}

\section{Introduction}

There is growing public interest in the decarbonization of the transportation sector, which today contributes almost 30\% of U.S. greenhouse gas emissions \cite{epa_emissions}, nearly 25\% of which can be attributed to medium-duty and heavy-duty vehicles (HDVs) \cite{transportation_emissions}. Of particular interest is the potential impact of the electrification of HDVs, specifically Class 7, 8 and 9 vehicles exceeding 26,000 pounds \cite{afdc}. %A recent study argues for the economic feasibility of such vehicles \cite{feasibility_HDEV}, however it does not address on their impact on the physical power grid.
There are approximately 275,000 HDVs (not necessarily electric) registered just in the state of Texas, which comprise 7\% of the nationwide total of 3.91 million Class 8 heavy vehicles\cite{fhwa, ata}.  Each HDV travels on average over 62,000 miles annually - nearly 5.5 times the distance traveled by a typical passenger car \cite{afdc}. The significance of electrification of such a fleet is amplified given the average heavy-duty electric vehicle (HDEV) is expected to draw anywhere between 75 kW and 600 kW while charging \cite{kW_1, kW_2, kW_3}. Any mass electrification of such large vehicles requires a substantial restructuring and shift from fuel-based to electricity-based charging infrastructure. This would potentially disrupt both electric grid operators and external market participants, and it is imperative for policy makers to adopt a data-driven, scientific approach to reconciling the market adoption of HDEVs with the realities of existing grid infrastructure.

With the rapid pace of EV adoption and nascent stage of regulatory oversight in this space, there have been numerous studies focused on the impact and coordination of light-duty vehicles \cite{muratori, Alizadeh_Rossi, nrel}. However, relatively little research has been focused on understanding the impact of rapid heavy-duty vehicle electrification on grid reliability. One recent Texas-based case study \cite{borlaug2021heavy} argues that most existing distribution-level substations have the capacity to accommodate significant levels of HDEV penetration without need for upgrades; however, the study does not consider the impact of large-scale HDEV electrification on critical electric grid parameters such as voltage violations at the transmission and distribution grid levels. %Preliminary studies hint at the possibility of system disruption from fast charging technology \cite{nrel}. %While the trend toward vehicle electrification is widely acknowledged in the public domain, substantive questions remain regarding the viability of rapid introduction. 

%The emergence of electric vehicles (EVs) as an alternative mode of transportation in the United States has prompted widespread speculation regarding their utility relative to incumbent internal combustion technology. 
% The electrification of heavy-duty vehicles is a nascent and rapidly emerging avenue for decarbonization of the transportation sector.  

In this paper, we examine the impacts of increased vehicle electrification on the electric grid infrastructure, with particular focus on HDEVs.
%present a novel method for analyzing the physical and economic impacts of increased vehicle electrification, with a particular focus on heavy-duty electric vehicles (HDEVs). 
We utilize a synthetic representation of the 2000-bus Texas transmission grid\cite{texas_2000}, realistic representations of multiple distribution grids in Travis county, Texas\cite{travis_county}, and transit data pertaining to HDEVs, to 
%as the starting point for our simulation frameworkLeveraging insights from geographic, power system and transit system data, we \cite{texas_2000}\cite{travis_county}
uncover the consequences of HDEV electrification, and expose the limitations imposed by the existing grid infrastructure. 

Our analysis reveals that grid-wide voltage problems spatio-temporally correlated with the mobility of HDEVs may occur even at modest penetration levels. In fact, we find that as little as 30000 HDEVs (11\% of the HDVs in Texas) charging simultaneously 
%in congested areas of the grid 
can destabilize the transmission grid. We also find that voltage violations at the distribution level can occur on the order of just tens of HDEVs charging simultaneously. This translates to just less than 5 commercial supercharging stations being fully occupied, with each vehicle charging at 75 kW to 600 kW \cite{kW_1,kW_2,kW_3}. 
%dependencies of  on the introduction of HDEVs.
Our analysis is generalizable over any set of transmission and distribution networks, and is not restricted to those in Texas, thus revealing the fundamental limitations of the underlying electric grid infrastructure that must be addressed before any serious HDEV electrification effort.

\section{Impact of Heavy-duty Vehicle Electrification on Transmission and Distribution Grids}
Our goal is to analyze the impact of HDEV integration on the reliability of transmission and distribution grids. In particular, we will focus on one critical grid parameter, namely grid voltage violations. We begin by modeling a heavy-duty electric vehicle (HDEV)  as a a heavy-duty vehicle which runs on electric storage, where each vehicle is associated with a state of charge (SOC), which decreases only when the vehicle moves, and increases only when the vehicle is charging. We assume that HDEVs consume an average of 2 kWh per mile \cite{kW_2}, and travel on highway roads at an average speed of 60 miles per hour. In contrast to smaller electric vehicles, HDEVs not only require larger amounts of energy (kWh) to fully charge, but also demand more charging power (kW) to reasonably serve the vehicles. HDEVs are expected to draw anywhere between 75 kW and 600 kW while charging \cite{kW_1, kW_2, kW_3}. However, in our simulations, we adopt a conservative median charge rate of 150 kW.

Each HDEV has direct impact on the electric power grid only when it is charging at an electric vehicle charging station (EVCS). While there may be several EVCS's connected to a single distribution grid, we assume that no HDEV charges at at more than one EVCS within the the same distribution grid in the short run, since HDVs travel long distances after each charge. Thus, when modelling the impact of HDEVs on distribution grids, it suffices to capture only their local power consumption, but not their movement within the neighborhood itself. Conversely, since EVCS's are geographically dispersed across multiple locations in the transmission grid (state-wide or wider), we believe that it is necessary to consider the movement of a fleet of vehicles within each transmission grid's encompassing geographic region to capture their spatiotemporal impact on critical grid parameters such as grid voltages.

\begin{figure*}[!h]
	\centering
	\includegraphics[width=0.85\textwidth]{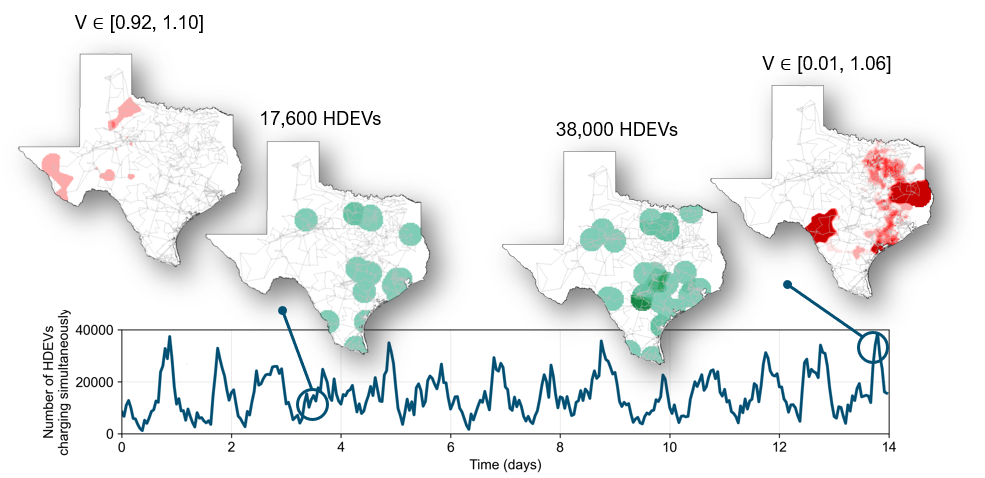}
	\caption{Spatiotemporal distribution of HDEVs over a 2-week period. Voltage range (pu) is shown above Texas maps. Voltage collapse is observed for 38,000 HDEVs charging simultaneously. Darker red in voltage contour plots refers to larger voltage magnitude deviations from nominal. Darker green refers to larger density of charging HDEVs.}
	\label{timeseries}
\end{figure*}

To model the movement of vehicles within the state of Texas, we first model the transportation network as a graph whose nodes represents the charging station locations and whose edges represent the roads along which any fleet of HDEVs must travel to relocate from one node to another. Edges are weighted by the average distance required to traverse from either node to the other (where each edge connects two nodes). Such a graph can be constructed and saved in the simulation ahead of time. Using such a graph, and the desired destinations and schedules of each HDEV fleet, we can obtain a spatiotemporally reasonable model of the movement of HDEV fleets. Based on such a model, we can simulate how many HDEVs there are at each EVCS, which in turn dictates how much power each node in the transmission network would have to serve. For example, in Fig. \ref{timeseries}, we observe that that the number of HDEVs charging simultaneously follows an oscillatory pattern, with the spatial distribution of those vehicles in the maps shown in green.

Based on these spatiotemporally realistic HDEV power consumption profiles, we simulate their impact on the transmission and distribution grids by observing the voltage deviation from the nominal value, which is normalized to 1 per unit. We use Python to interface with PowerWorld for transmission level computations, and to interface with OpenDSS for distribution level computations. Given HDEV power consumption at each bus, voltage is observed and visualized geographically as a critical factor in quantifying the impact of HDEVs. For an in-depth description of the modeling and simulation procedure, including the modeling of the transmission and distribution grids, the charging stations, the vehicles and the road network, we refer the reader to the Supplementary Material. 

For the transmission grid, we carried out these simulations on a synthetic representation of the 2000-bus Texas transmission grid\cite{texas_2000}. Defining the voltage deviation as per unit values outside the range of $[0.95, 1.05]$, Fig. \ref{timeseries} shows that it takes about 17,000 HDEVs charging simultaneously to cause minor yet non-negligible voltage deviations at multiple locations in the transmission grid, while 38,000 HDEVs charging simultaneously lead to much more severe deviations, including voltage collapse, as shown in the figure (voltage close to zero). Note that the figure shows a per unit voltage range of $[0.01, 1.06]$. The 0.01 value refers to voltage collapse in simulation. However, in a real-world setting, the corresponding amount of vehicles wouldn't have been able to have been served due to protection equipment preventing this before the occurence of voltage collapse.

\begin{figure*}[!h]
	\centering
	\includegraphics[width=0.99\textwidth]{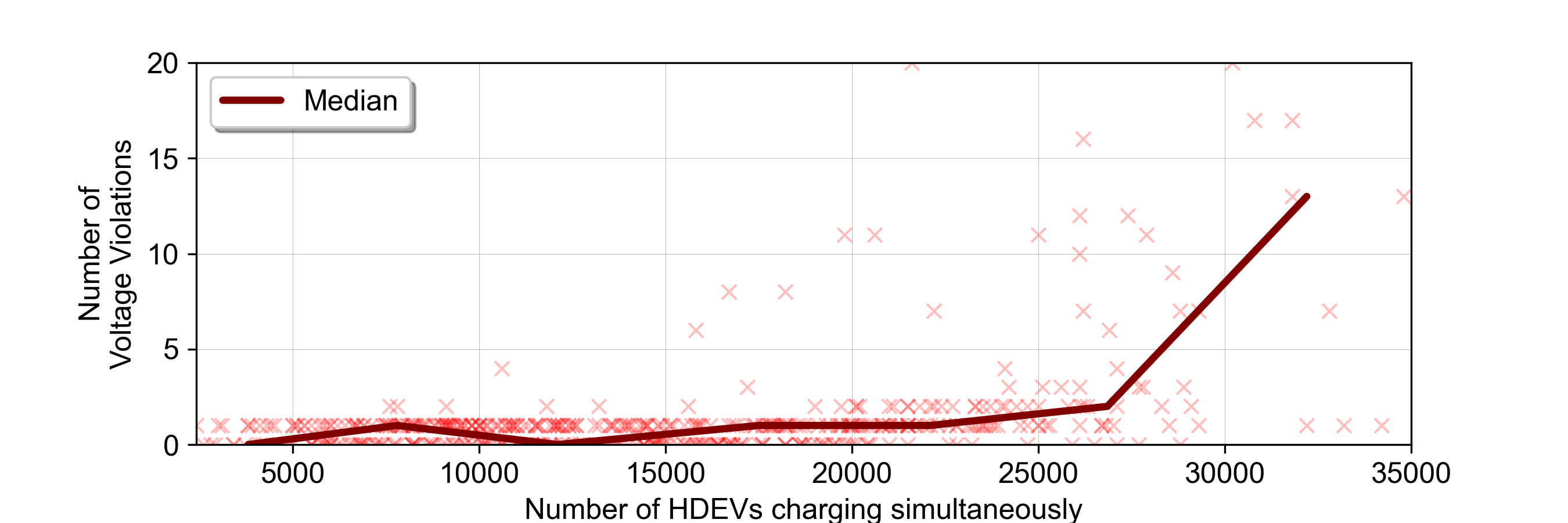}
	\caption{Number of voltage violations vs HDEVs charging simultaneously, sampled from multiple simulations such as those in Fig. \ref{timeseries}. Median of experiments is shown in solid line. Results show accelerating median impact at approximately 30,000 HDEVs.} % How I chose 30000 as a reasonable number of HDEVs charging simultaneously (and cite it ... 273,000 heavy-duty vehicles).
	\label{nVV_vs_nHDEV}
\end{figure*}

Fig. \ref{nVV_vs_nHDEV} examines the median number of buses with voltage violations based on the number of HDEVs charging simultaneously. % to severely impact the transmission grid. 
% We show median values, rather than worst-case, in  Fig. \ref{nVV_vs_nHDEV}. 
These results demonstrate that at around 30,000 HDEVs, there is a sudden increase in median number of buses with voltage violations. While this median number is on the order of tens of buses, the worst-case is on the order of hundreds of buses, such as what is shown in Fig. \ref{timeseries} with the voltage collapse. We refer the reader to the Supplementary Material for a comparison between the median and the worst case scenarios.

\begin{figure*}[!h]
	\centering
	\includegraphics[width=0.95\textwidth]{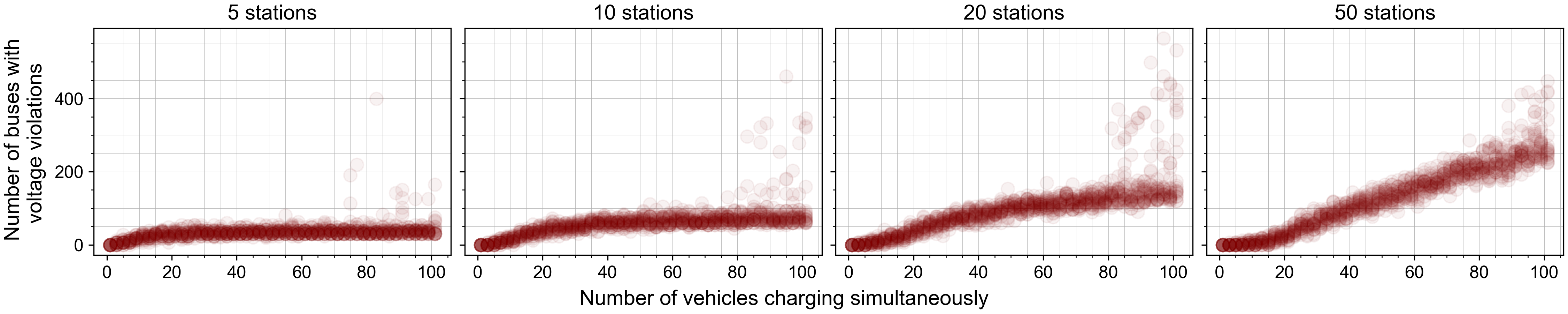}
	\includegraphics[width=0.95\textwidth]{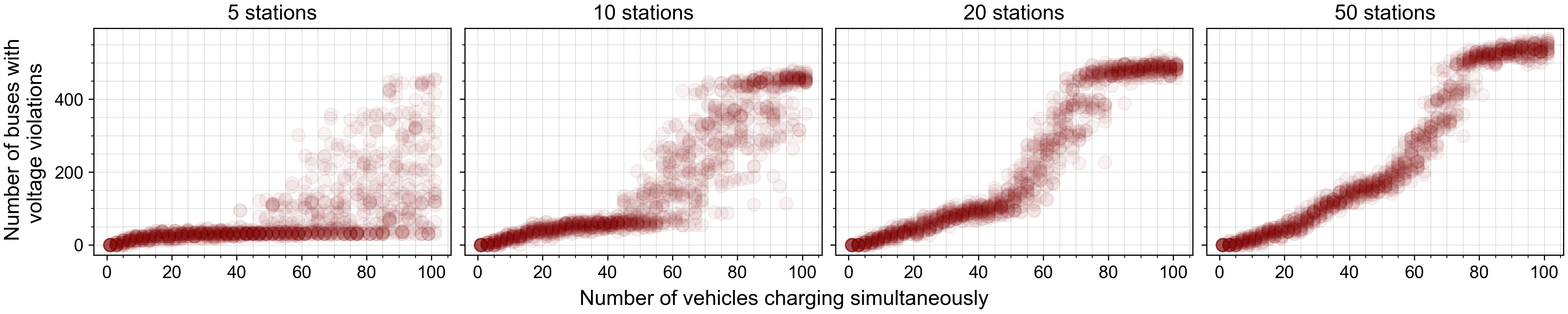}
	\includegraphics[width=0.95\textwidth]{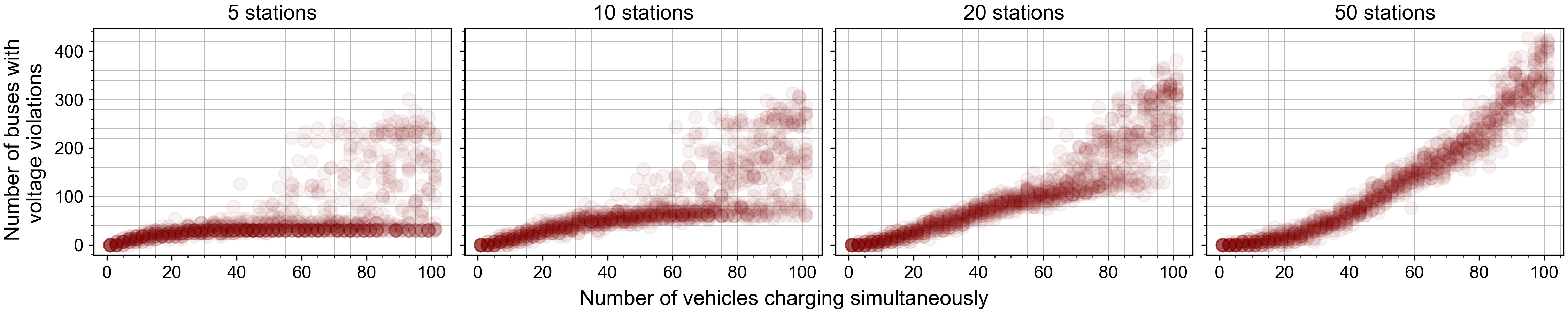}
	\includegraphics[width=0.95\textwidth]{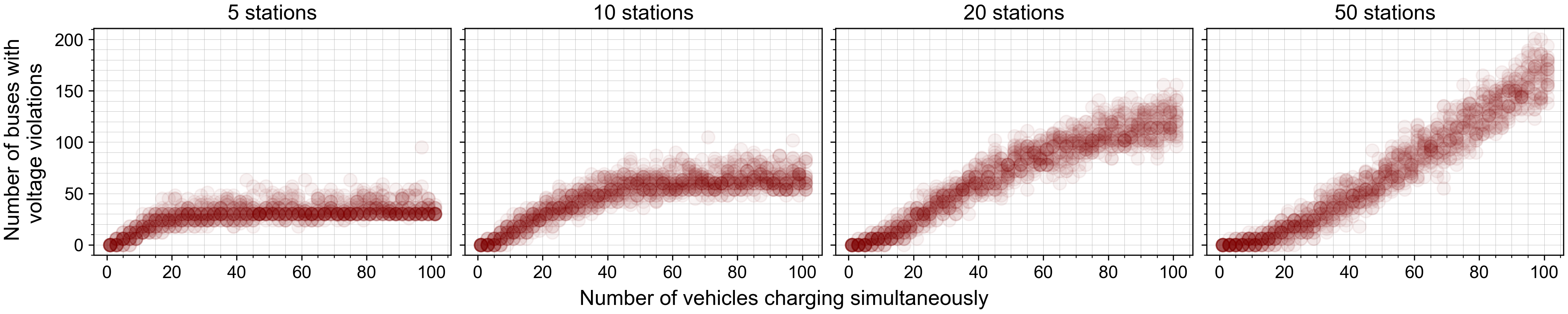}
	\caption{For four independent distribution grid models (one per row), and different number of stations per grid (each column), the effect of heavy-duty vehicle electrification on voltage violations is shown. Counter-intuitively, the more stations used, the more number of voltage violations there are. We refer the reader to the Supplementary Material for a more exhaustive set of simulation results.}
	\label{violations_distribution}
\end{figure*}

We then examine the  impact of HDEVs on distribution grids. Fig. \ref{violations_distribution} reveals an additional dimension (beyond just the number of HDEVs charging simultaneously), which is the number of charging stations (EVCSs) that are available to serve power. Each row in the figure represents a distinct and independent distribution grid, all based on feeders within Travis county, Texas \cite{travis_county}. The results in Fig. \ref{violations_distribution} show that lowering the number of charging stations, and providing more ports per station, could prove to be less impactful on local distribution grids in terms of decreasing the number of buses with voltage violations. An intuitive justification for this is that a heavy concentration of load locally leads to voltage violations most expressed locally.

% Use end of this section (and Fig. 3) to transition to Discussion & Conclusion (Policy Implications).

% See this paper (Siva, Kiyeob, Xinbo) for ideas on supplementary material: https://www.sciencedirect.com/science/article/pii/S258900422101693X#tbl1

\section{Discussion and Conclusions}

We simulate the impact of heavy-duty electric vehicles (HDEVs) on the electricity infrastructure by considering a synthetic Texas-based transmission and distribution grid. We find that it takes around 30,000 HDEVs, corresponding to electrification of only $\approx$ 11\% of the HDVs in Texas, to severely impact the transmission grid, and on the order of merely tens of vehicles per local distribution grid to create a serious under-voltage problems for electric utilities. We capture the impact of such vehicles by identifying the number and location of buses with voltage violations, which can guide investment strategies aimed at ensuring grid voltage reliability.

Our findings reveal to policymakers the potential adverse impacts of rapid electrification of HDVs on grid reliability, and provide insights on how to determine suitable locations for infrastructure upgrade investments. In particular, our case study indicates that upgrading local distribution grids will need to be top priority. While upgrading infrastructure will be crucial, smarter scheduling of HDEVs through spatial and temporal redistribution could serve to alleviate costs and increase social welfare. In summary, this work motivates the need for serious consideration of grid impacts for any successful HDV electrification effort. 

% Discussions should be brief and focused. In some disciplines use of Discussion or `Conclusion' is interchangeable. It is not mandatory to use both. Some journals prefer a section `Results and Discussion' followed by a section `Conclusion'. Please refer to Journal-level guidance for any specific requirements.

% Conclusions may be used to restate your hypothesis or research question, restate your major findings, explain the relevance and the added value of your work, highlight any limitations of your study, describe future directions for research and recommendations.

%%%%%%%%%%%%%%%%%%%%%%%%%%%%%%%%%%%%%%%%%%%%%%%%
% END STUFF
%%%%%%%%%%%%%%%%%%%%%%%%%%%%%%%%%%%%%%%%%%%%%%%%

\Urlmuskip=0mu plus 1mu\relax  % split url
\newpage
\bibliography{ref}

%%%%%%%%%%%%%%%%%%%%%%%%%%%%%%%%%%%%%%%%%%%%%%%%
% SUPPLEMENTARY MATERIAL
%%%%%%%%%%%%%%%%%%%%%%%%%%%%%%%%%%%%%%%%%%%%%%%%
\newpage
{\raggedright\sffamily\bfseries\fontsize{20}{25}\selectfont Supplementary Material}
\vspace{1em}

\section*{\uppercase{Key Resources and Availability}}

% \subsection*{Lead Contact}
% Further information and requests for resources and materials should be directed to and will be fulfilled by the Lead Contact, Dr. Le Xie (le.xie@tamu.edu).

\subsection*{Simulation Tools}
Transmission grid power flow calculations were run using PowerWorld, and distribution grid power flow calculations were run using OpenDSS. Both PowerWorld and OpenDSS were interfaced with using Python 3.7 on a Windows 10 machine, which facilitates automated simulations.

\subsection*{Synthetic Grid Models}
The 2000-bus synthetic model of the Texas grid \cite{texas_2000} was used for transmission grid simulations. Distribution grid simulations were based on a set of models which use realistic data from feeders in Travis country, Texas \cite{travis_county}.

\subsection*{Load and Renewables Time-series}
Included in our GitHub repository \cite{our_github} is a set of time-series of year-long wind turbine power outputs, solar panel power outputs and load consumption at various buses in the transmission grid. Wind data was obtained using the System Advisor Model (SAM) \cite{nrel_SAM}, solar data was obtained using the National Solar Radiation Database \cite{nrel_solar} and load data is modified based on data from reference \cite{texas_2000}.

\subsection*{Heavy-Duty Electric Vehicle Parameters}
Physical parameters of HDEVs were based on data obtained from Fleet DNA by NREL \cite{fleet_dna} and from Tesla \cite{tesla_semi}. Such data includes average speed on highways, average energy consumed per mile, charging power and charging capacity.

\subsection*{Data and Code Availability}
Python code used in our simulation is open-source and can be found on GitHub repository \cite{our_github}. Data used is either referenced in the repository or directly attached there.

\section*{\uppercase{Method Details}}

\subsection*{Simulating Impact on Transmission Grid}
\begin{figure*}[!h]
	\centering
	\includegraphics[width=0.95\textwidth]{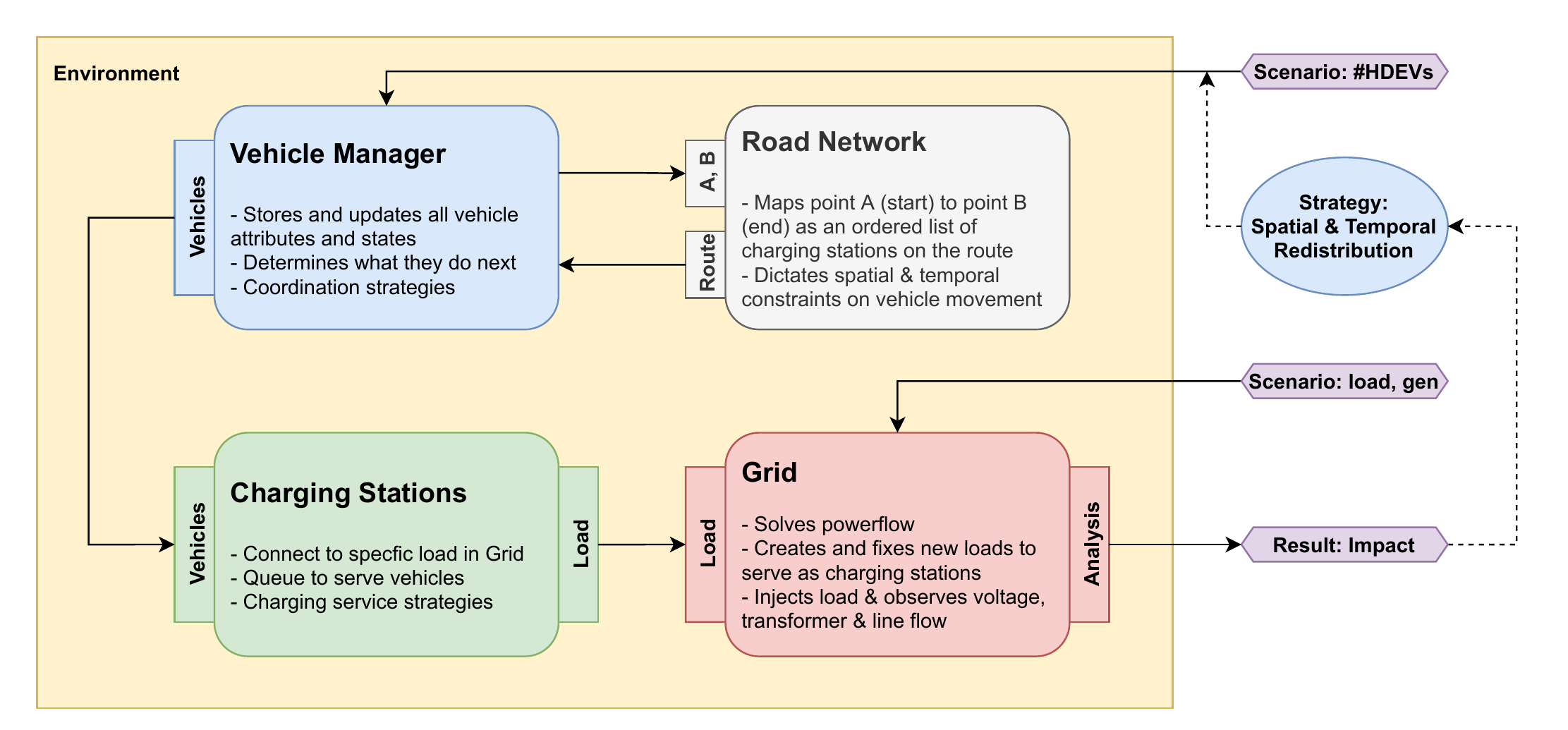}
	\caption{Flowchart of framework for simulating the impact of HDEVs on the transmission network. The user specifies scenarios, and the environment automatically generates the results accordingly.}
	\label{env_flowchart}
\end{figure*}

We divided the simulation process into 4 distinct modules, which are labelled in Fig. \ref{env_flowchart} as \textbf{Grid}, \textbf{Charging Stations}, \textbf{Vehicle Manager} and \textbf{Road Network}. The detailed descriptions of the modules are as follows:
\begin{itemize}
    \item The \textbf{Grid} module computes power flow results given the load consumption, solar power output and wind power output across the grid. It can do so with or without the presence of HDEVs in the system. For the purposes of this study, we rely on a Python-based interface with PowerWorld, using SimAuto \cite{ESA}, for both solving such computations and for obtaining a geographical visual representation of bus voltage magnitude corresponding to the results.
    \item The \textbf{Charging Stations} module is responsible for connecting HDEVs to the power grid. Simply put, each station is aware of how many vehicles are currently charging locally, it aggregates their net power consumption, and it signals the \textbf{Grid} module to consume said consumption at the appropriate bus. This module is also responsible for sorting and distributing the vehicles across the charging ports based on pre-defined strategies. Station locations are pre-determined.
    \item The \textbf{Vehicle Manager} module accounts for the state of all HDEVs present in the system, including their location (on the road or at a station), battery charge level, and desired schedule and destinations. At any point in time, this module communicates with the \textbf{Charging Stations} module where each vehicle is so that the latter knows which vehicles demand charging service and where. For each vehicle, this module communicates with the \textbf{Road Network} module to estimate travel constraints and decide which routes to assign to the vehicle, and when to stop and recharge.
    \item The \textbf{Road Network} module serves the \textbf{Vehicle Manager} as mentioned above. It relies on a constant weighted undirected graph, whose nodes represent charging stations and whose edges are weighted by the distance required for a vehicle to traverse from either end of the edge to the other. This is a simplified model of the road network. Potential expansions include considering a time-varying directed graph, influenced by real-time traffic congestion.
\end{itemize}

In this simulation environment, the user specifies which period of the year to consider. This is used to initialize load and renewable time-series for the duration of the simulation. For example, results shown in the main paper reflect a two-week period in summer. Then, the user specifies the \textit{initial number of HDEVs} in the system and the \textit{arrival rate} of HDEVs into the system. At the start of the simulation, each vehicle is automatically assigned a \textit{starting location} and a \textit{desired destination}. Once the vehicle reaches its destination, it \textit{departs} from the system. This arrival and departure process was adopted to enable a less memory-intensive procedure than the alternative of fixing a large number of HDEVs in the system from the start.

Once the simulation is complete, we observe a record of how many vehicles there were at each point in time which were: \textit{charging simultaneously}, \textit{idling at the stations}, and \textit{moving on the road}. We tune the aforementioned \textit{arrival rate} to minimize the number of vehicles idling at the stations. The simulation also records a time-series of the impact on the grid expressed in terms of voltage deviations from some nominal value. Once we know both the temporal variation of voltage, and given the locations of measurements, we obtain a spatiotemporally realistic estimate of the impact of HDEVs on the power grid.

In Fig. \ref{env_flowchart}, the bubble titled \textbf{Strategy: Spatial \& Temporal Redistribution} indicates that vehicles need not behave in an open-loop fashion, whereby the only deciding factor in their charging and routing policies is their desired destination. Vehicles could also adopt a closed-loop grid-aware approach based on observing set of timely metrics, such as nodal electricity pricing and physical grid impact per charge, and deciding accordingly.

\subsubsection*{Median vs. Worst-Case impact}
In Fig. \ref{timeseries} of the main paper, we plot the median impact of HDEVs to show how most of the buses are affected by increasing penetration of such vehicles. In Fig. \ref{hist_worst_median_nVV} below, we illustrate that even though the median may be on the order of tens, the worst case may be on the order of hundreds. That is, even if only a few buses are affected \textit{most} of the time, occasionally much more buses are. The results in Fig. \ref{hist_worst_median_nVV} show that for a two-week period where the median is 26 buses with voltage violations, the worst-case is 386. Table \ref{table:num_HDEVS} below summarizes the number of HDEVs for that same simulation period.

\begin{figure*}[!h]
	\centering
	\includegraphics[width=0.9\textwidth]{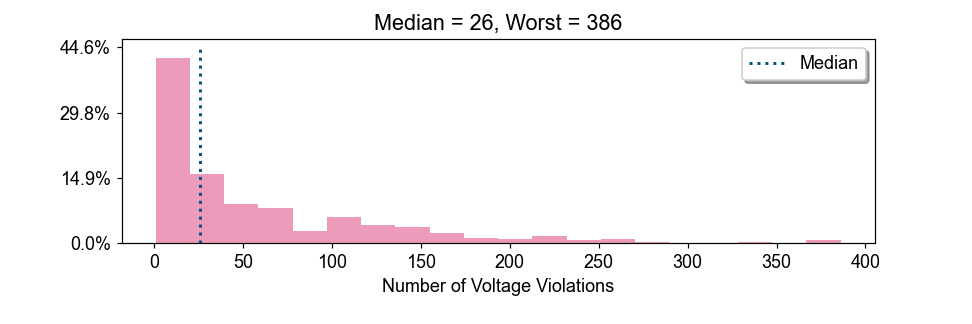}
	\caption{Number of buses with voltage violations for a two-week simulation period, where the corresponding number of vehicles is described in Table \ref{table:num_HDEVS}.}
	\label{hist_worst_median_nVV}
\end{figure*}

\begin{table}[h]
\centering
\begin{tabular}{ccccc}
\hline
Minimum & Maximum & Median & Mean & Standard Deviation\\
\hline
1400 & 40800 & 13100 & 14588 & 8226  \\
\hline
\end{tabular}
\caption{Statistics on the number of HDEVs for the two-week period referenced in Fig. \ref{hist_worst_median_nVV}.}
\label{table:num_HDEVS}
\end{table}

\subsection*{Simulating Impact on Distribution Grid}
In contrast with the simulations related to the transmission grid (illustrated in Fig. \ref{env_flowchart}), when simulating the impact of HDEVs on distribution grids, we do not consider a road network for the following reason. Since HDVs are long-haul, they do not perform frequent stops within the same neighborhood. Rather, between every charge and the next, they tend to travel to more distant charging stations. Hence, each vehicle's impact on any given distribution grid is temporary and singular in time. For this reason, we deem it sufficient to simply consider the net impact of a collection of HDEVs at a single snapshot of time. That is, the only parameters left to consider when simulating the impact of such vehicles on distribution grids are (a) the total number of vehicles charging simultaneously in a given grid, and (b) the spatial distribution of those vehicles across the grid.

In Fig. \ref{violations_distribution_6} (and the subsequent figures continuing it), we show an exhaustive set of simulations over 36 independent distribution grids, based on data from Travis county \cite{travis_county}. Each grid is represented by a row in the figures. The horizontal axis represents the first parameter (number of HDEVs), but to search over all possible spatial distributions is computationally intractable. We reduce the search space by first considering how many possible stations there are in the grid. Each column in the figures represent one of 5, 10, 20 or 50 stations. Then, for each case, we uniformly sample the location of stations and number of vehicles plugged per station. This yields a large number of outcomes, each represented by a semi-transparent circle on the plot. More densely colored portions of the plots indicate that many of the scenarios yielded similar results.

We observe the following pattern from the results. Less charging stations, and more charging ports per station, yields less impact on distribution grids for the same number of vehicles charging simultaneously.

\begin{figure*}[!h]
	\centering
	\includegraphics[width=0.95\textwidth]{images/p1uhs0_1247--p1udt2142.png}
	\includegraphics[width=0.95\textwidth]{images/p1uhs10_1247--p1udt22980.png}
	\includegraphics[width=0.95\textwidth]{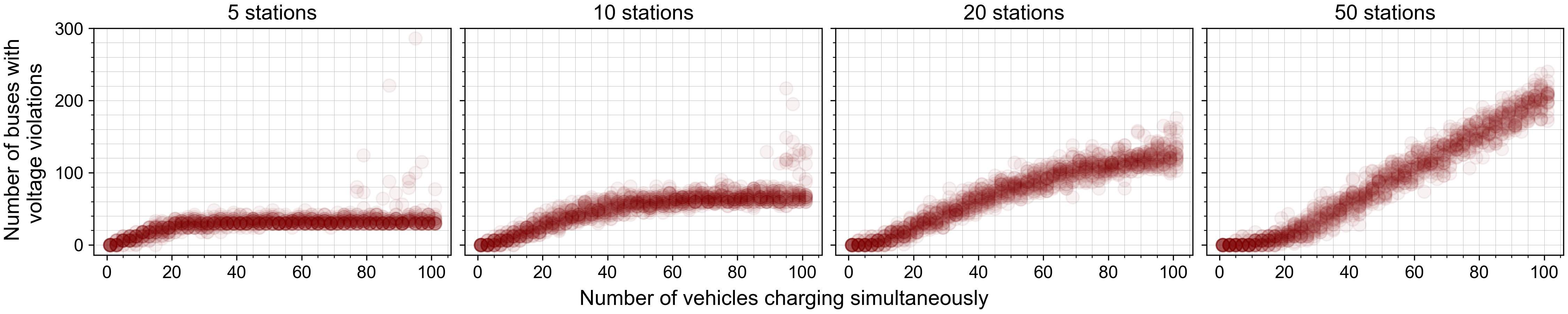}
	\includegraphics[width=0.95\textwidth]{images/p1uhs14_1247--p1udt21168.png}
	\includegraphics[width=0.95\textwidth]{images/p1uhs14_1247--p1udt9467.png}
	\includegraphics[width=0.95\textwidth]{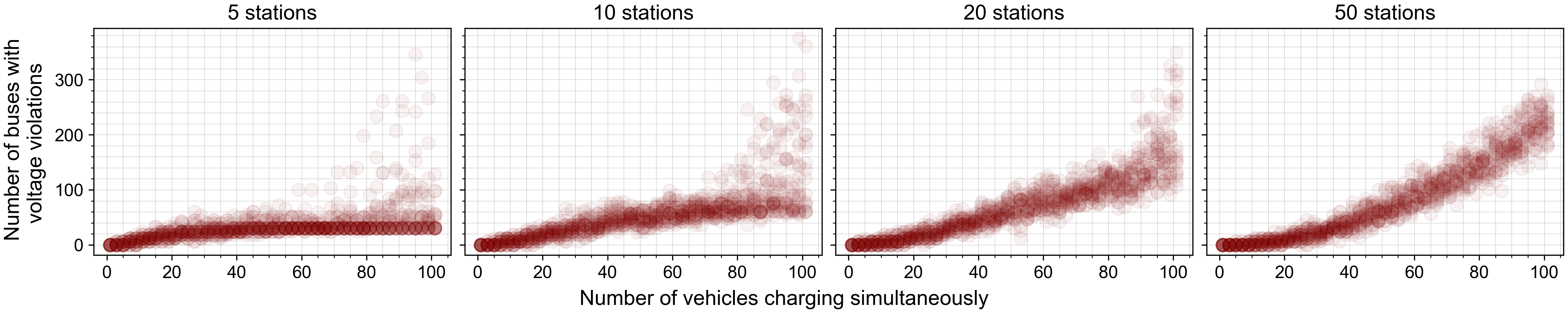}
	\caption{For 6 independent distribution grid models (one per row), and different number of stations per grid (each column), the effect of heavy-duty vehicle electrification on voltage violations is shown. Counter-intuitively, the more stations used, the more number of voltage violations there are.}
	\label{violations_distribution_6}
\end{figure*}
\addtocounter{figure}{-1}

\begin{figure*}[!h]
	\centering
	\includegraphics[width=0.95\textwidth]{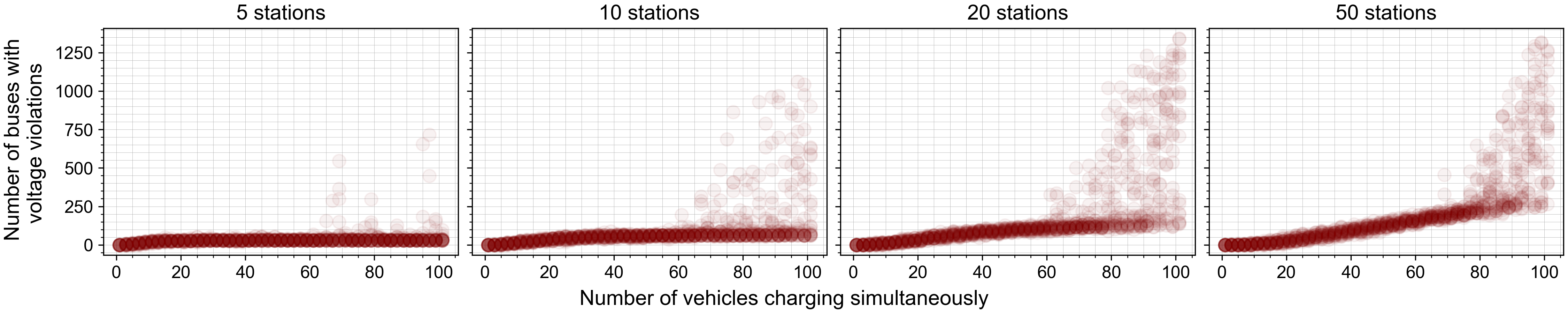}
	\includegraphics[width=0.95\textwidth]{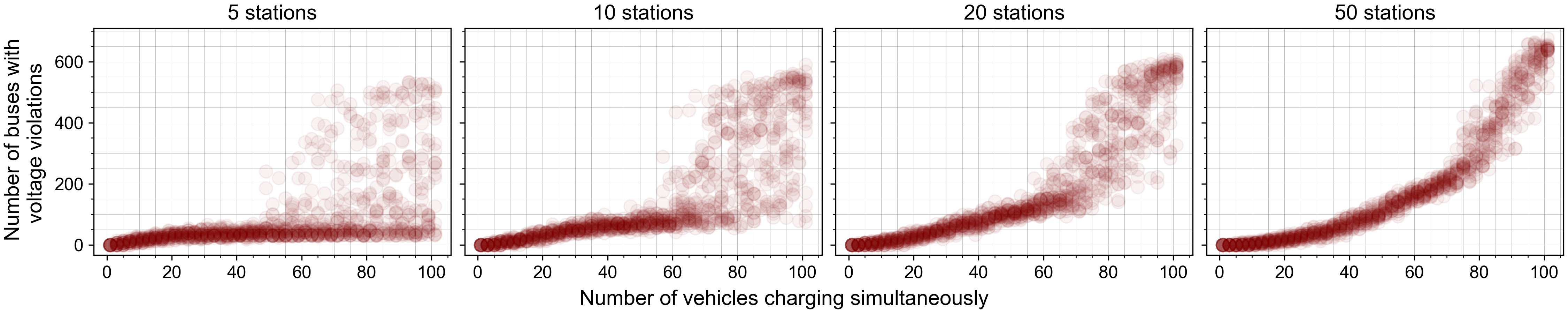}
	\includegraphics[width=0.95\textwidth]{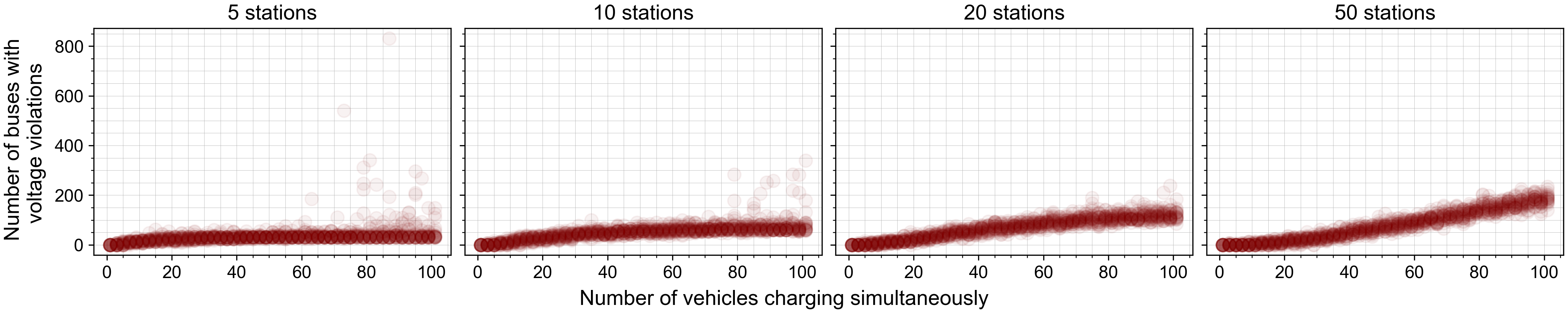}
	\includegraphics[width=0.95\textwidth]{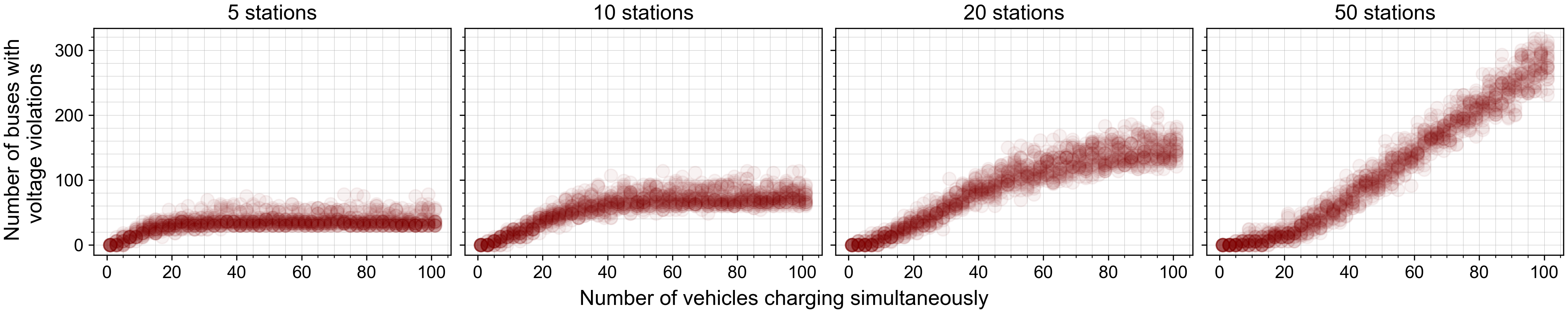}
	\includegraphics[width=0.95\textwidth]{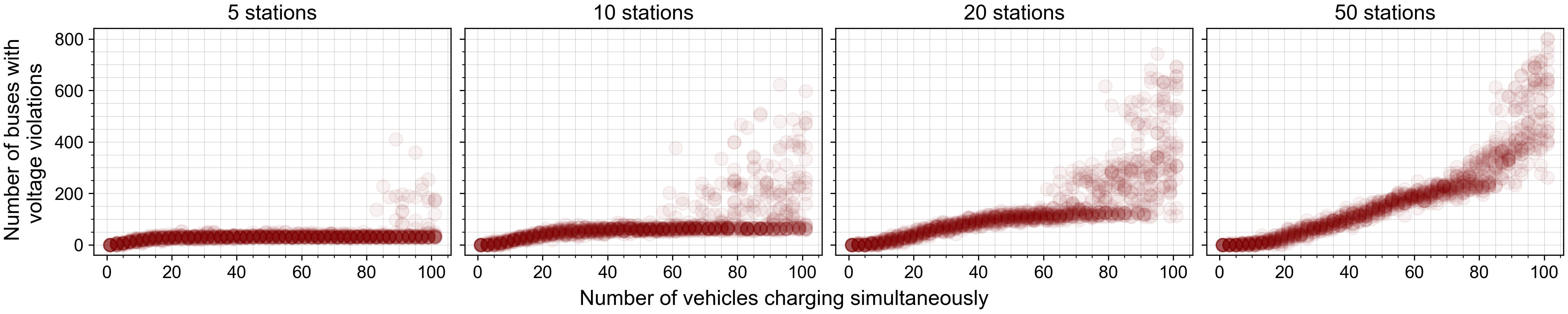}
	\includegraphics[width=0.95\textwidth]{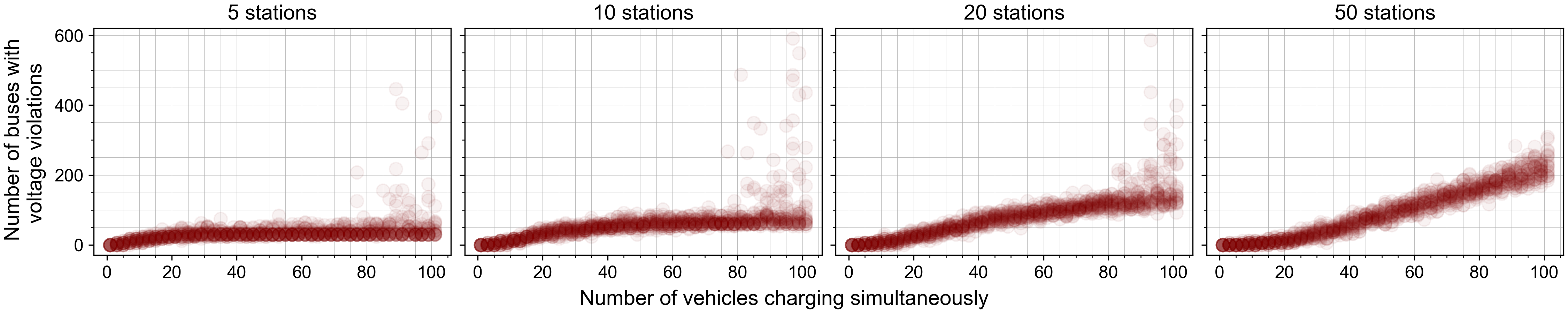}
	\caption{Continued.}
\end{figure*}

\addtocounter{figure}{-1}
\begin{figure*}[!h]
	\centering
	\includegraphics[width=0.95\textwidth]{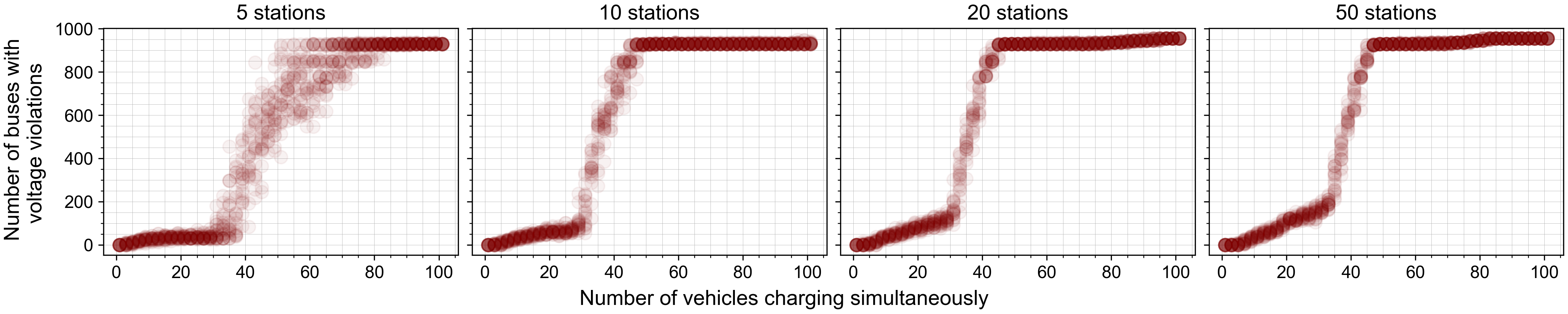}
	\includegraphics[width=0.95\textwidth]{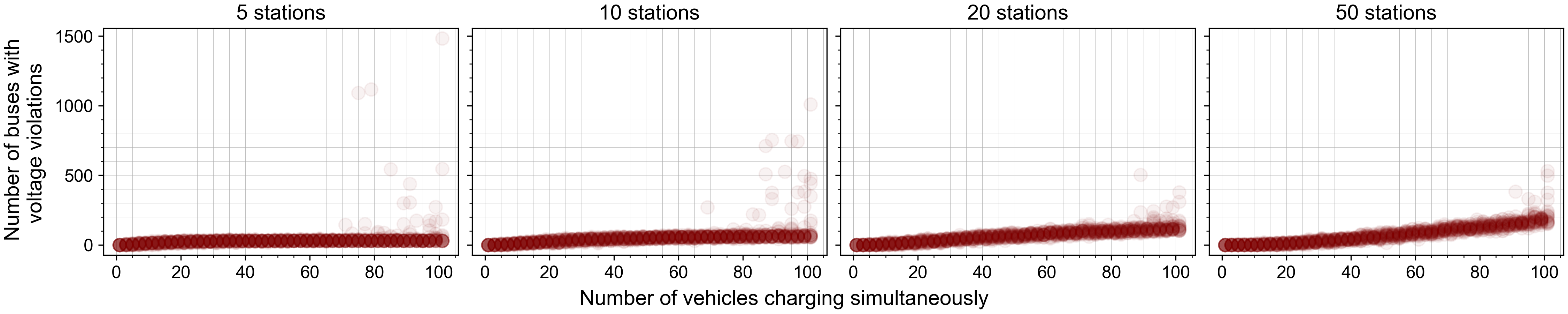}
	\includegraphics[width=0.95\textwidth]{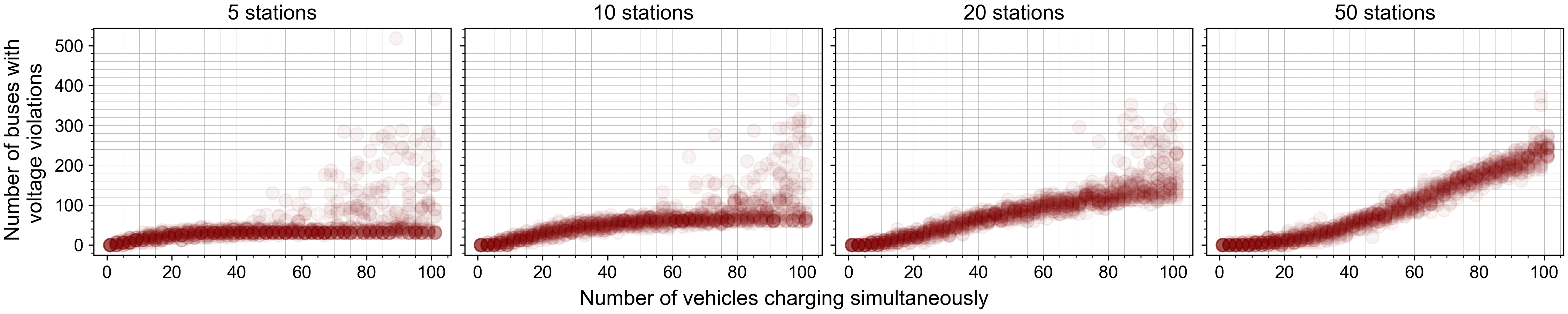}
	\includegraphics[width=0.95\textwidth]{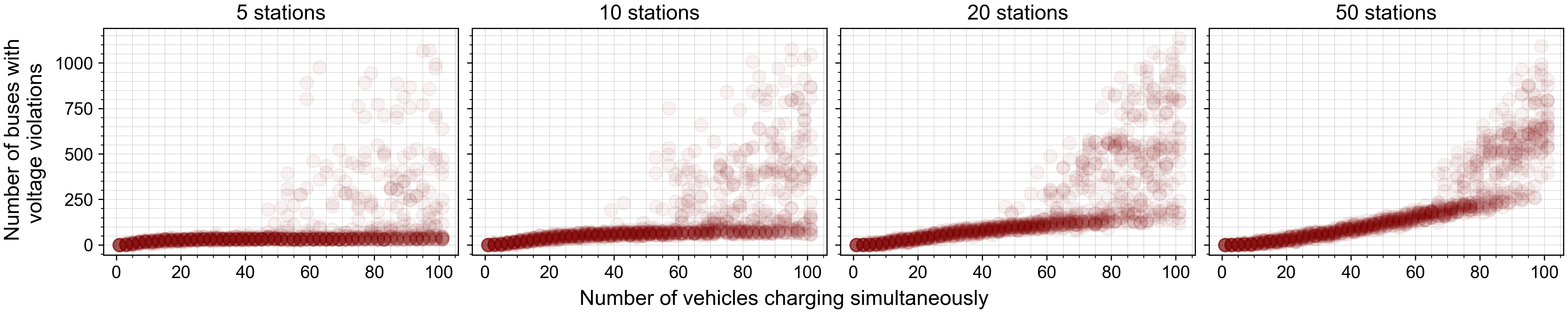}
	\includegraphics[width=0.95\textwidth]{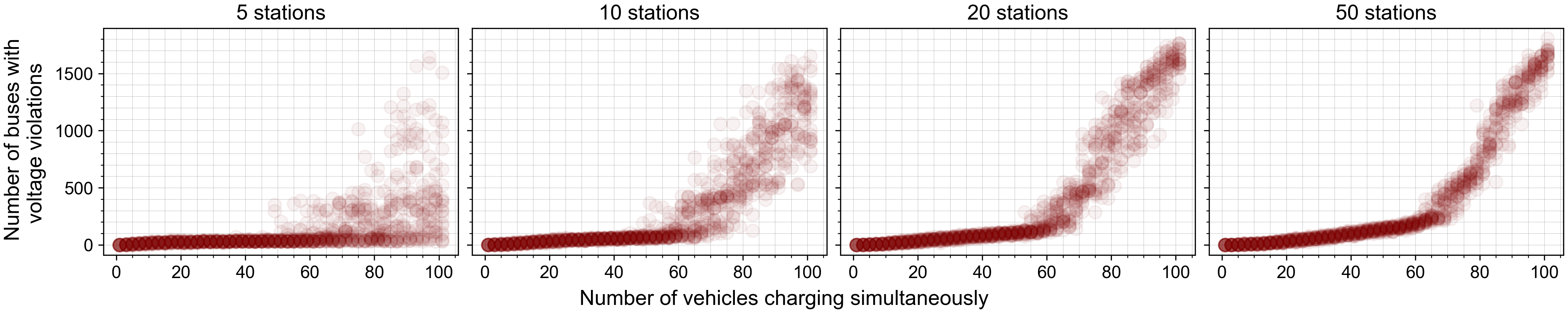}
	\includegraphics[width=0.95\textwidth]{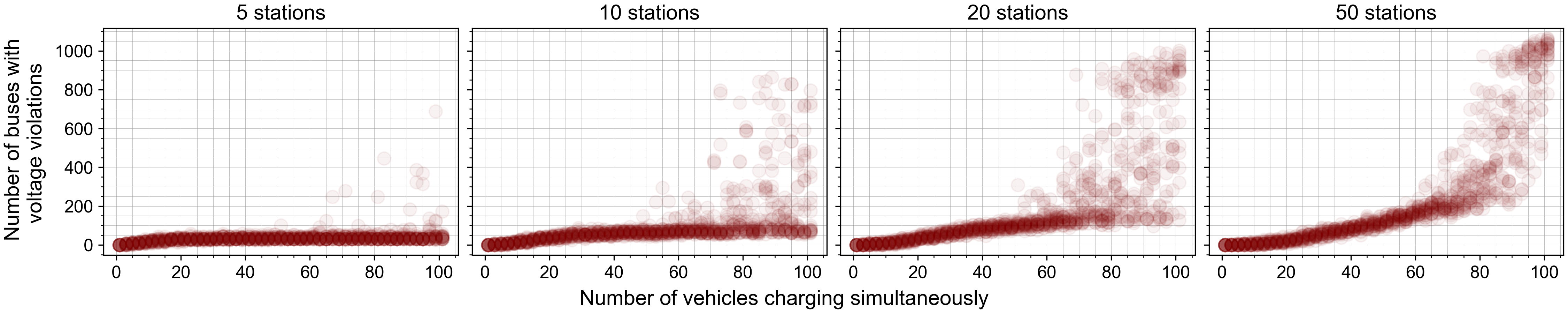}
	\caption{Continued.}
\end{figure*}

\addtocounter{figure}{-1}
\begin{figure*}[!h]
	\centering
	\includegraphics[width=0.95\textwidth]{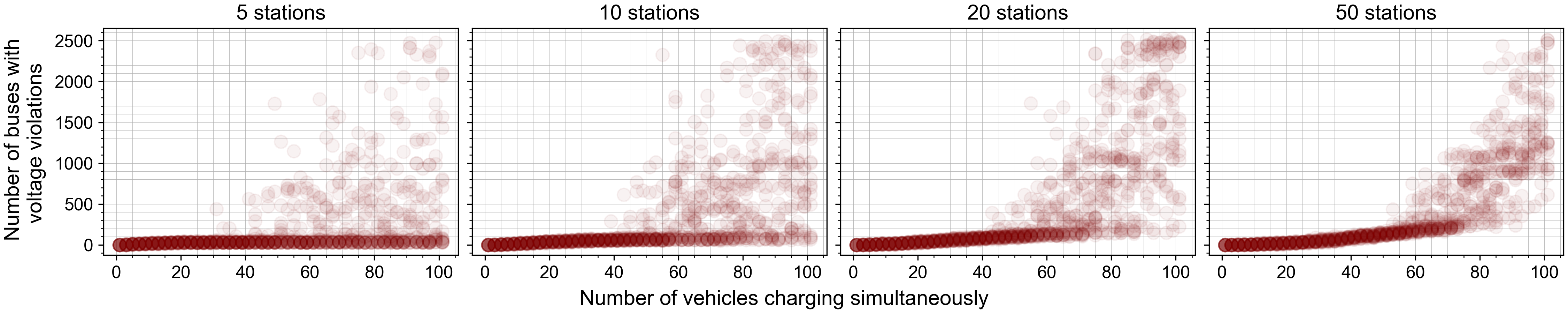}
	\includegraphics[width=0.95\textwidth]{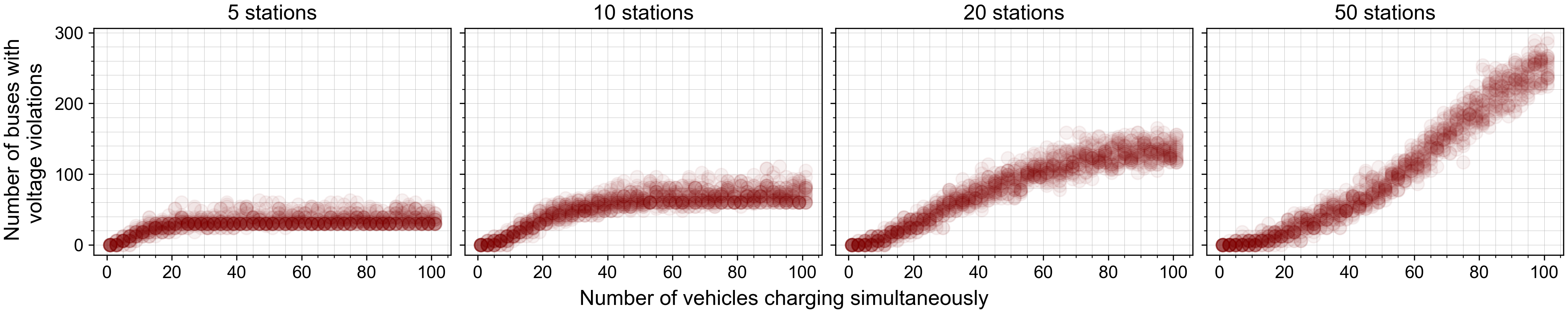}
	\includegraphics[width=0.95\textwidth]{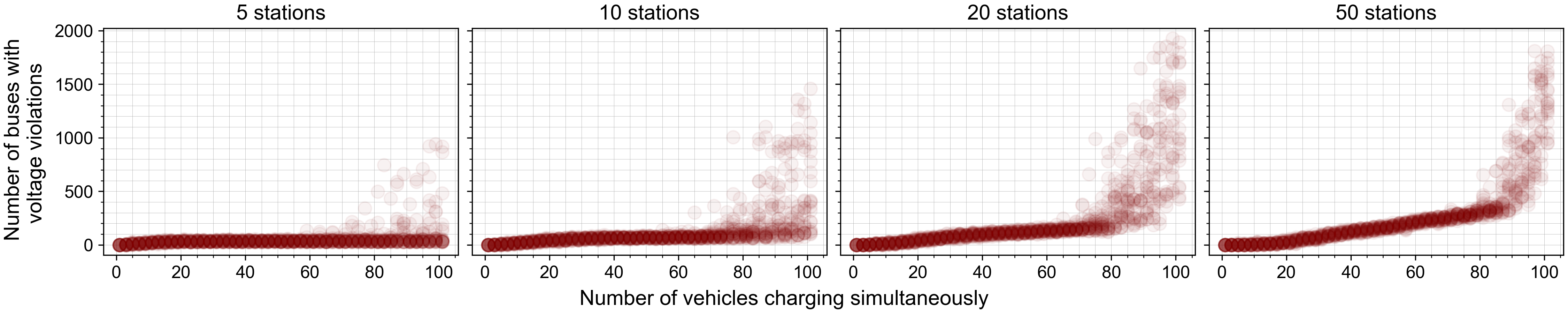}
	\includegraphics[width=0.95\textwidth]{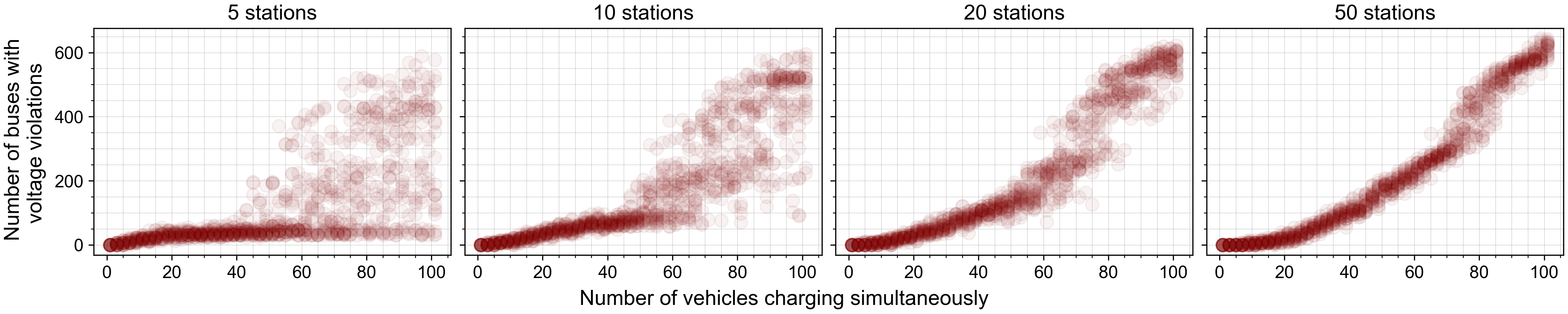}
	\includegraphics[width=0.95\textwidth]{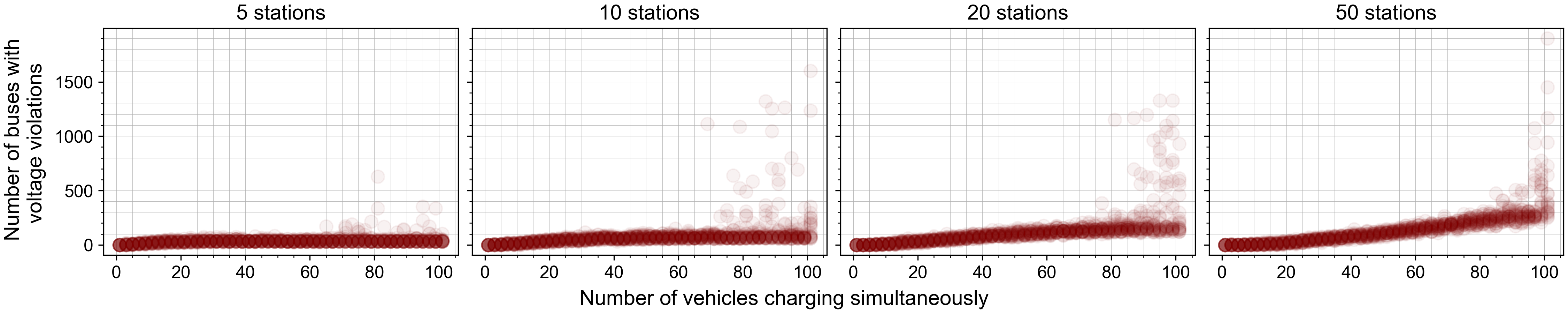}
	\includegraphics[width=0.95\textwidth]{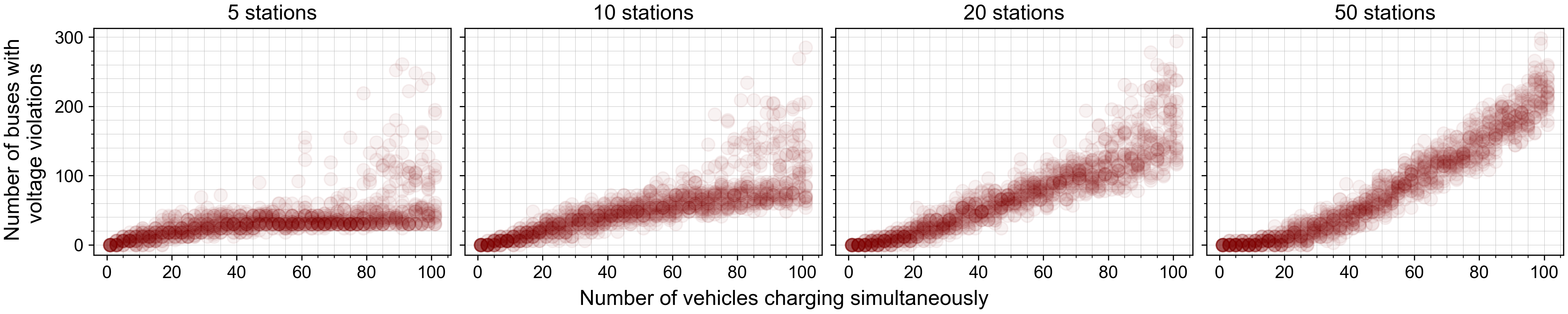}
	\caption{Continued.}
\end{figure*}

\addtocounter{figure}{-1}
\begin{figure*}[!h]
	\centering
	\includegraphics[width=0.95\textwidth]{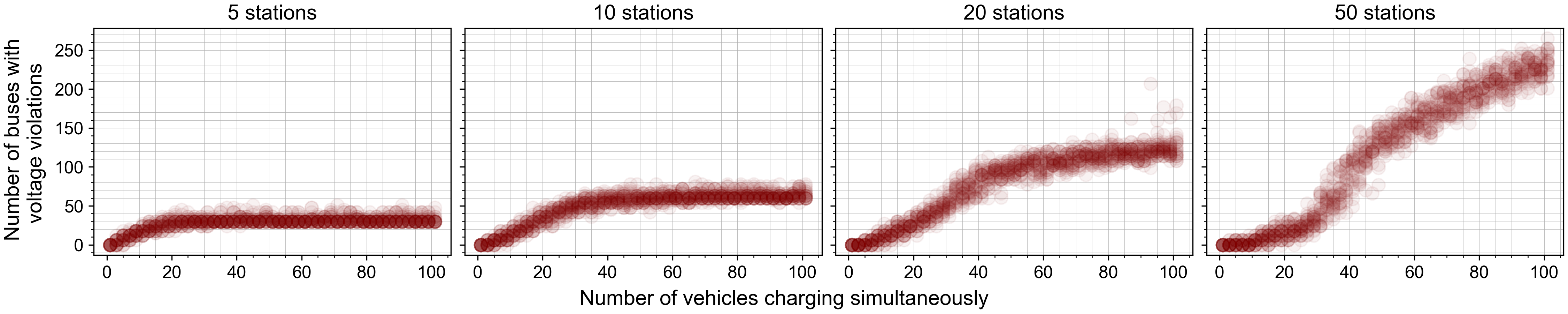}
	\includegraphics[width=0.95\textwidth]{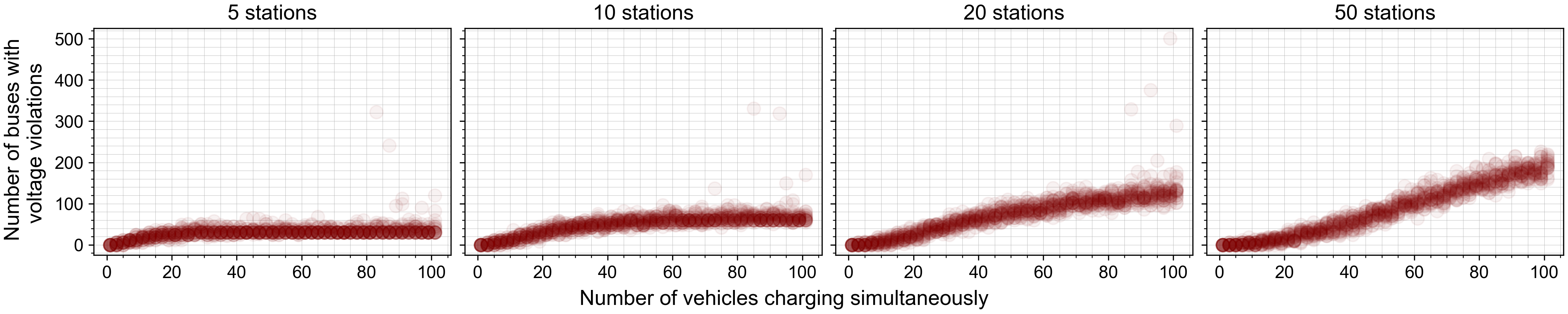}
	\includegraphics[width=0.95\textwidth]{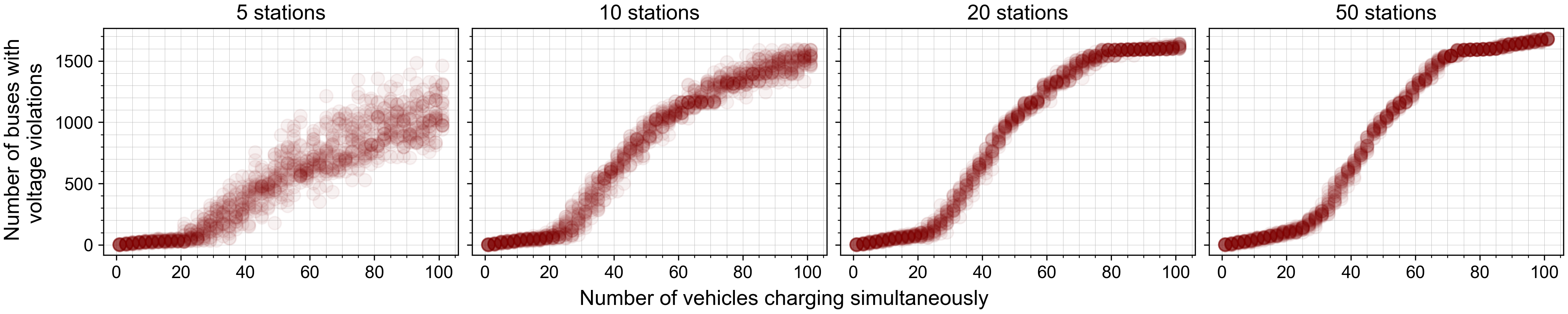}
	\includegraphics[width=0.95\textwidth]{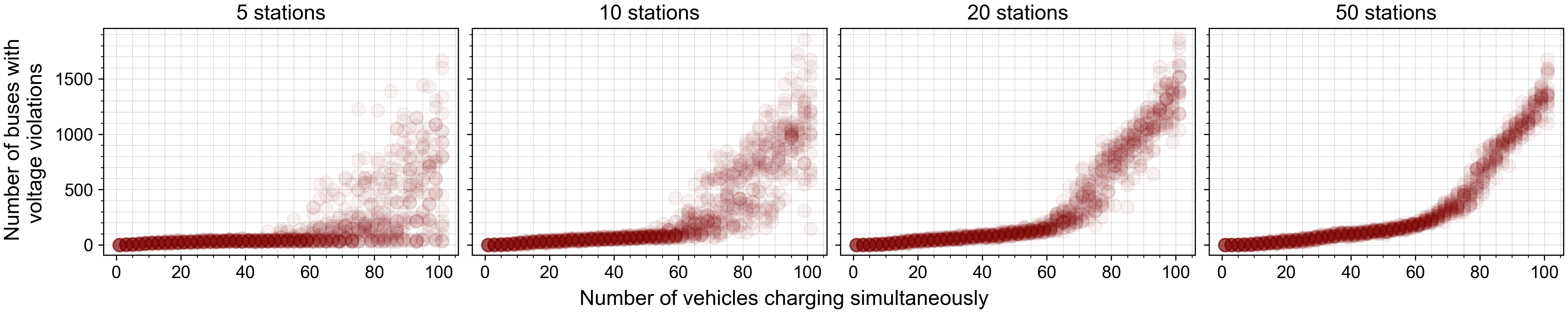}
	\includegraphics[width=0.95\textwidth]{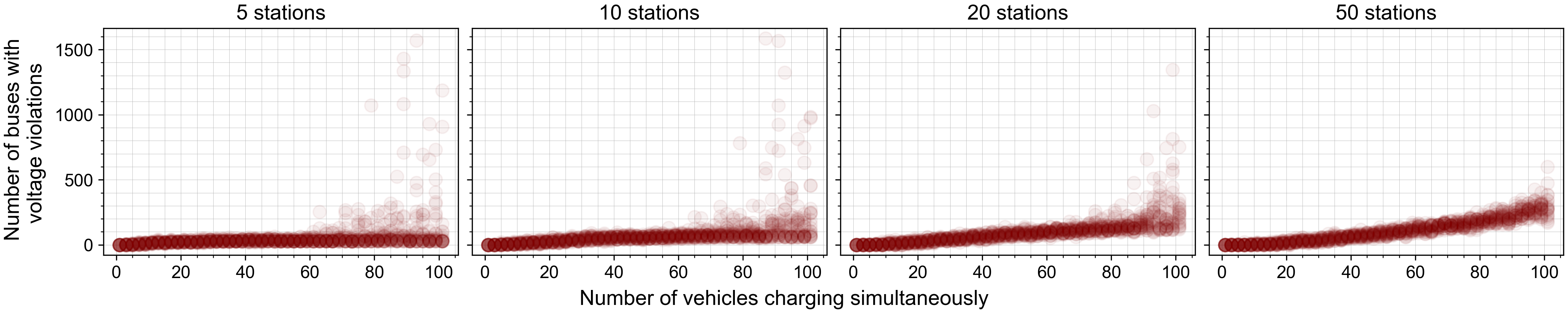}
	\includegraphics[width=0.95\textwidth]{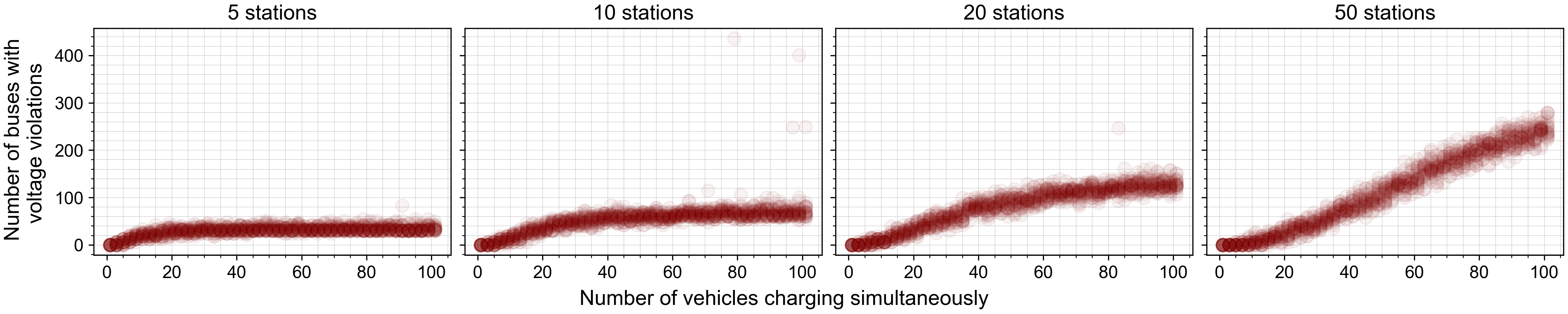}
	\caption{Continued.}
\end{figure*}

\addtocounter{figure}{-1}
\begin{figure*}[!h]
	\centering
	\includegraphics[width=0.95\textwidth]{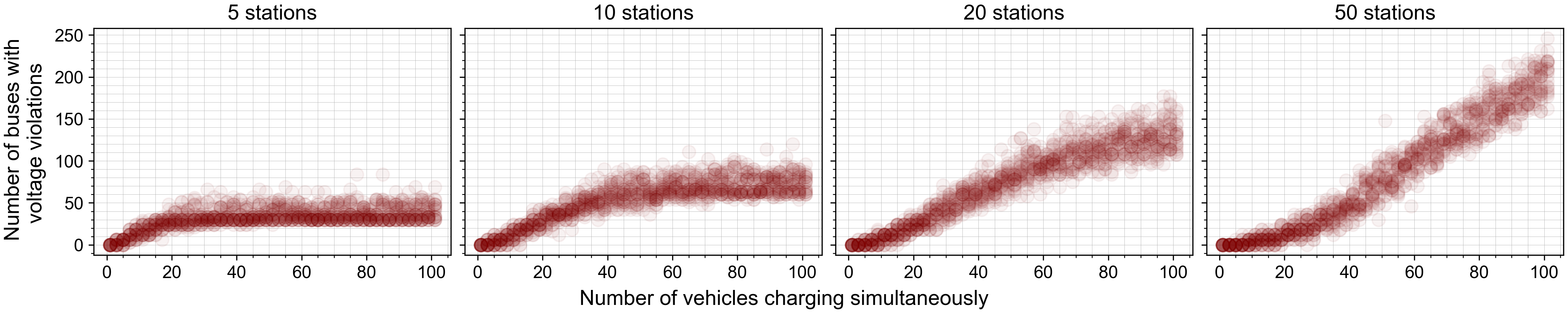}
	\includegraphics[width=0.95\textwidth]{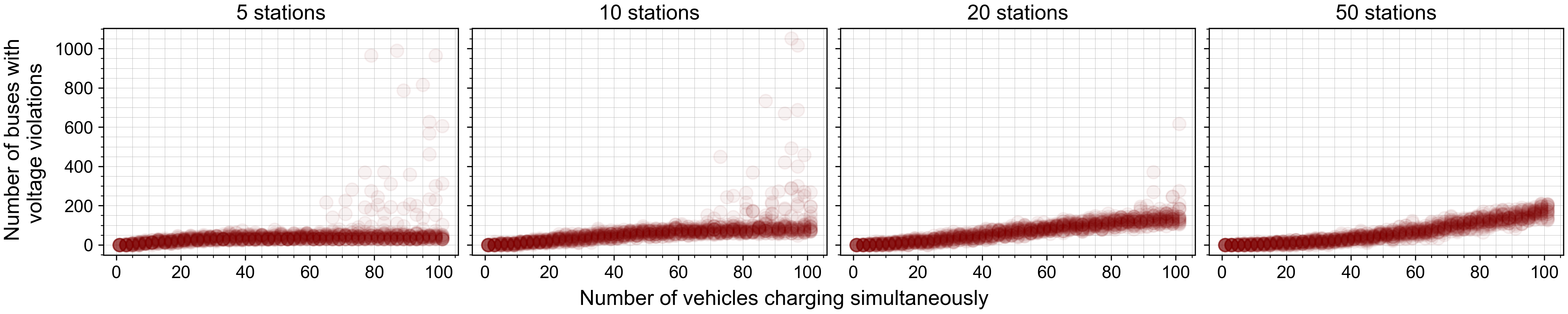}
	\includegraphics[width=0.95\textwidth]{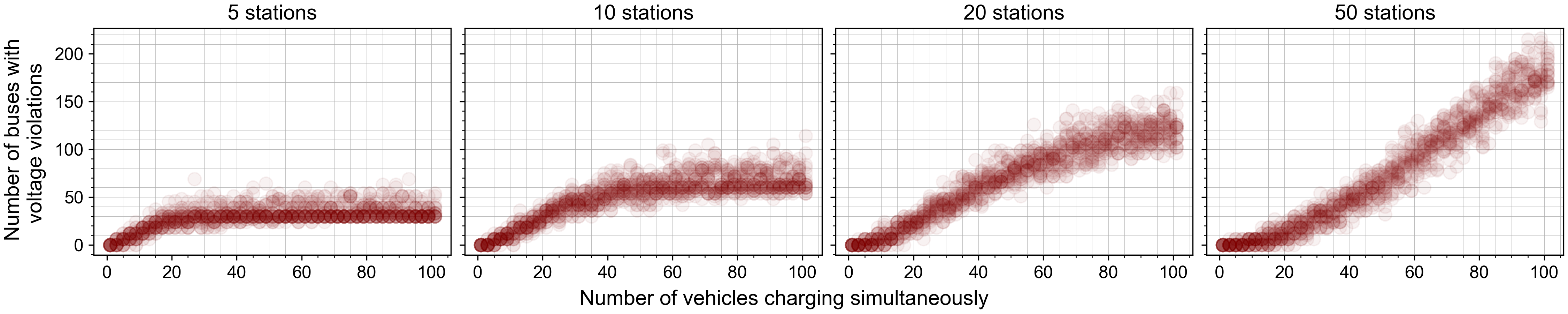}
	\includegraphics[width=0.95\textwidth]{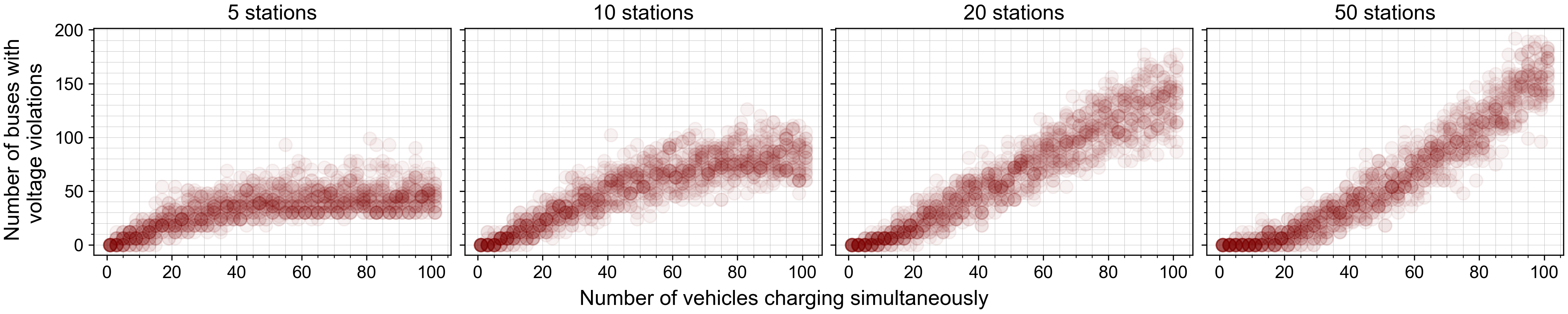}
	\includegraphics[width=0.95\textwidth]{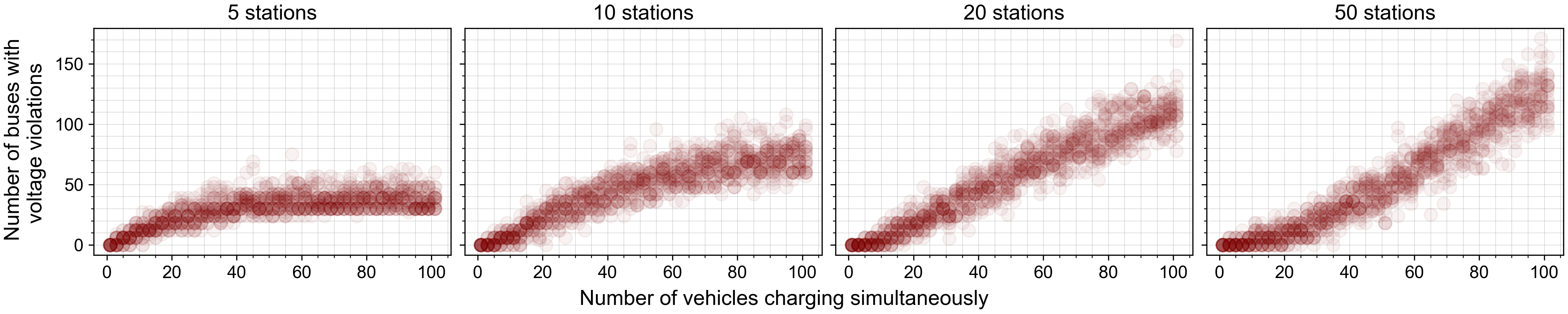}
	\includegraphics[width=0.95\textwidth]{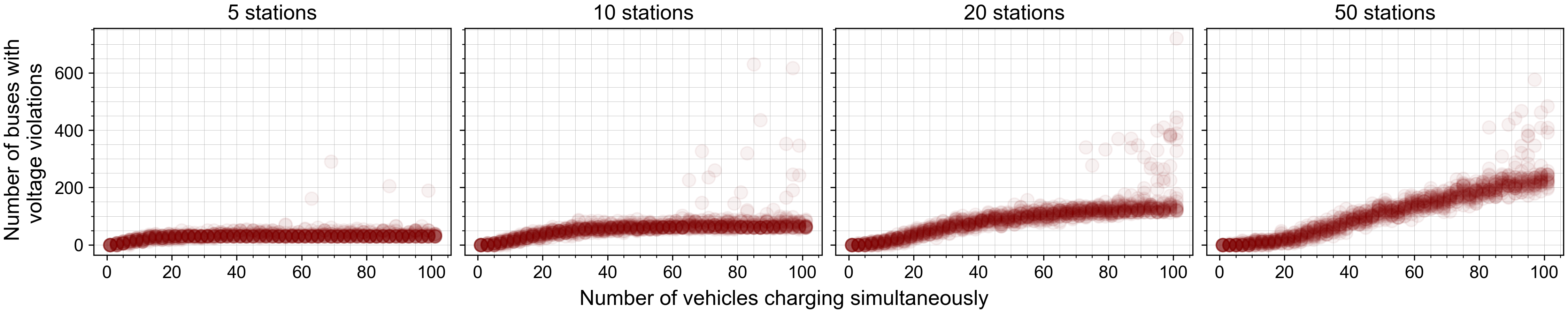}
	\caption{Continued.}
\end{figure*}

\end{document}